\documentclass[hyper,letterpaper]{JHEP} 

\JHEPspecialurl{http://jhep.sissa.it/JOURNAL/JHEP3.tar.gz}

\usepackage{epsfig,multicol,amssymb,amsmath}

\newcommand\fverb{\setbox\pippobox=\hbox\bgroup\verb}
\newcommand\fverbdo{\egroup\medskip\noindent%
            \fbox{\unhbox\pippobox}\ }
\newcommand\fverbit{\egroup\item[\fbox{\unhbox\pippobox}]}

\newenvironment{myindentpar}[1]%
{\begin{list}{}%
         {\setlength{\leftmargin}{#1}}%
         \item[]%
}
{\end{list}}
\newbox\pippobox

\title{Gravity and Electroweak Symmetry Breaking in a RSI/RSII Hybrid Model}

\author{Brian Glover$^a$ and Jong Anly Tan$^a$\\
        $a$ Particle Theory Group, Department of Physics,
College of William and Mary, Williamsburg, VA 23187-8795 \\

    E-mail: \email{baglov@wm.edu}, \email{jmtanx@wm.edu}}

\abstract{We present a hybrid RSI/RSII model in which we both solve the hierarchy problem and produce a continuum of KK graviton modes.  In this model, four dimensional gravity can be reproduced and the radion mode can be stabilized.  We then modify the hybrid gravity model to include SU$(2)_{\rm{L}} $xSU$(2)_{\rm{R}} $xU$(1)_{\rm{B-L}}$ bulk gauge fields.  Electroweak symmetry is broken by the choice of appropriate boundary conditions.  By adjusting the size of one region of the extra dimension, we show that the $S$ parameter can be decreased while protecting the $\rho$ parameter from corrections.  We find that as the $S$ parameter is decreased by $\sim 60\%$, $M_{Z'}$ and $M_{W'}$ stay below 1800 GeV, protecting unitarity.}


\begin{document}

\section{Introduction}\label{sec:intro}
The idea of warped extra dimensions was first introduced in 1983 when Rubakov and Shaposhnikov suggested that a vanishing 4D cosmological constant would result if a 5D bulk vacuum energy was tuned to cancel the large 4D vacuum energy of the Standard Model (SM) fields \cite{Rubakov}.  This work was popularized in 1999 when Randall and Sundrum introduced two famous examples of warped extra dimensions which led to interesting and distinct phenomenology (hereafter called RSI \cite{RSI} and RSII \cite{RSII}).  In the first model (RSI), a finite warped extra dimension living between a positive and a negative tension brane was used to solve the hierarchy problem.  This model predicts Kaluza-Klein (KK) graviton excitations to have masses on the order of a few TeV which could possibly be detected at the Large Hadron Collider (LHC) in the near future.  In the RSII model, Randall and Sundrum considered a warped infinite extra dimension.  Although they no longer solved the hierarchy problem, they found that four dimensional gravity can still be reproduced in an infinite extra dimension since the corrections to Newton's Law at large distances are suppressed on the positive tension brane.

\begin{figure}[h]
\centering
\epsfig{figure=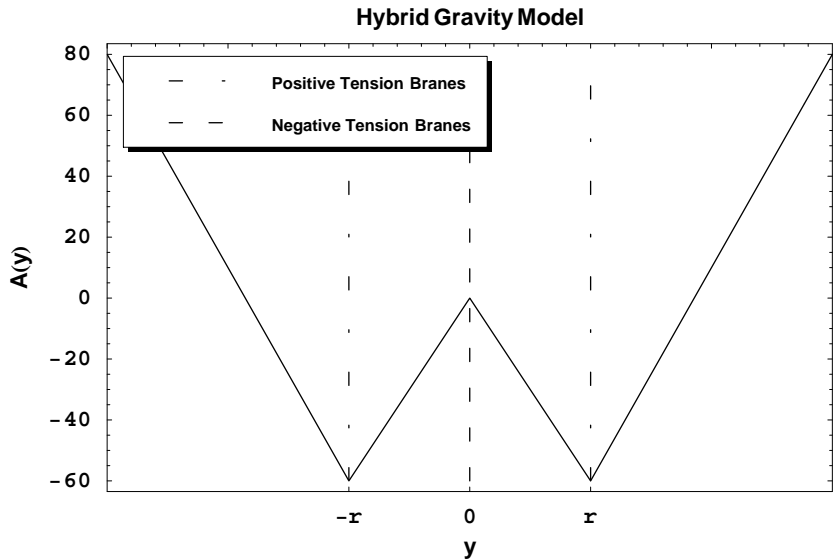,width=5.0in}
\caption{The Hybrid RSI/RSII gravity model.  The space is orbifolded around $y=0$ and extends to infinity.}
\label{fig:gravitymetric}
\end{figure}

Since these models were first introduced, many extensions of their work have been proposed.  Some of these extensions include adding extra branes to the bulk of RSII \cite{LR, GRS, Kogan}, localizing gravity on thick branes \cite{Thick Branes}, adding SM fields to the bulk of RSI \cite{BulkFields}, Higgless models in an RSI background \cite{Higgsless}, etc. In one of these models \cite{GRS}, an extra negative tension brane was included in the bulk of the infinite extra dimension of RSII.  This model, if stable, was designed to solve the hierarchy problem as in RSI but with an infinite extra dimension.  However, it was found that when the scalar gravity mode (radion) of the five dimensional graviton is carefully considered, the theory becomes unstable \cite{radionproblem}.  This instability arose since the kinetic term of the radion in these theories was found to be negative \cite{pilo}.  The bulk stress tensor violates the positivity of energy condition and the brane is unstable to crumpling.  More recently, Agashe et al. \cite{Agashe} pointed out that if one could stabilize a  IR-UV-IR model with Z$_2$ parity about the UV brane, one could address the hierarchy problem naturally.  They argue that in an alternate UV-IR-UV model, one would have to add large brane kinetic terms in order to solve the hierarchy problem.  In Section~\ref{sec:gravity} we propose a model in which the negative tension brane is placed at an orbifold fixed point with positive tension branes living in the bulk of an infinite, warped extra dimension (see Fig. \ref{fig:gravitymetric}).  The metric is given by $ds^2=e^{-A(y)}dx^2+dy^2$ where the warp factor is
\begin{equation}\label{eq:ymetric}
    A(y)=\left\{ \begin{array}{ll}
                    -2 k_1 |y| & \mbox{if $0 \leq |y| \leq r$}\\
                    2 k_2 |y| - 2(k_1+k_2)r & \mbox{if $|y| > r$}.\end{array} \right.
\end{equation}
As in Lykken and Randall \cite{LR}, this theory has a continuous KK spectrum while also solving the hierarchy problem.  However, the phenomenology of our model is more of a hybrid between RSI and RSII in which the KK gravitons of RSI become resonances.  Placing a negative tension brane at an orbifold fixed point projects out the negative energy mode of the radion and therefore allows the theory to be stabilized.  We calculate the gravitational spectrum and show how this theory can be stabilized.

\begin{figure}[h]
\centering
\epsfig{figure=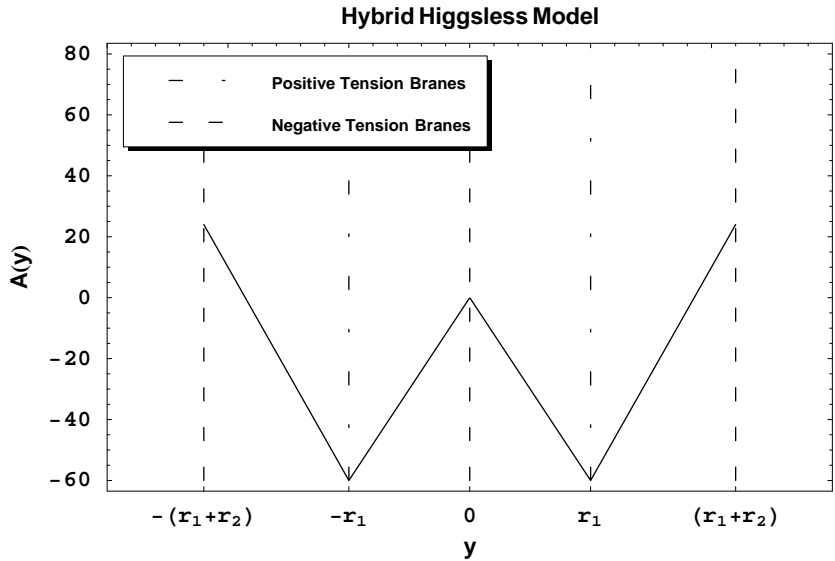,width=5.0in}
\caption{The Hybrid RSI/RSII higgsless model. The space is orbifolded around $y=0$ and ends at the location of the outside negative tension branes ($y=\pm (r_1+r_2)$).}
\label{fig:higgslessmetric}
\end{figure}

Warped extra dimensions have also proven to be interesting for models of Higgsless Electroweak Symmetry Breaking.  In Cacciapaglia et al. \cite{Higgsless}, SU$(2)_{\rm{L}} $xSU$(2)_{\rm{R}} $xU$(1)_{\rm{B-L}}$ gauge fields were included in the bulk of AdS space.  They showed that breaking SU$(2)_{\rm{R}} $x U$(1)_{\rm{B-L}}$ down to U$(1)_{\rm{Y}}$ on the Planck brane protects the $\rho$ parameter from corrections since the broken SU(2) gauge group shows up as a custodial symmetry in the holographic interpretation \cite{holographic}.  It was found, as in technicolor theories, that an order one S parameter is produced in conflict with experiments.  In order to address this problem, a Planck brane kinetic term was added which was found to decrease the $S$ parameter but at the price of destroying unitarity.  They also added a U$(1)_{B-L}$ brane kinetic term to the TeV brane which also lowered the $S$ parameter but at the price of making $T$ nonzero.  More recently Carone et al. \cite{Carone} showed that a holographic UV-IR-UV model can be constructed, with SU$(2)_{\rm{L}} $xU$(1)_{\rm{B-L}}$ gauge fields in the bulk, in which a custodial symmetry is generated without introducing a SU$(2)_{\rm{R}}$ gauge group.  They found that like the standard higgsless model, the S parameter is too large.  In Section~\ref{sec:gauge} we modify our hybrid model to include gauge fields in the warped extra dimension.  Following Cs\`{a}ki et al. \cite{Higgsless}, we include SU$(2)_{\rm{L}} $x SU$(2)_{\rm{R}} $x U$(1)_{\rm{B-L}}$ gauge fields in the bulk and use boundary conditions to break the symmetry in order to reproduce the SM on one of our branes.  In order to have a normalizable photon, we have brought in another negative tension brane from infinity to cut off the space at an orbifold fixed point (see Fig. \ref{fig:higgslessmetric}).  We find corrections to the $\rho$ parameter to be suppressed, signaling that an approximate custodial symmetry is preserved.  We calculate oblique corrections in this model and find that as the added slice of the extra dimension increases, the S parameter decreases.  We stress that this method of reducing the $S$ parameter appears to keep corrections to the $\rho$ parameter suppressed while preserving unitarity for a decrease in $S$ up to $60\%$.

\section{Gravity in the Hybrid Model}\label{sec:gravity}
Our theory is defined by placing a negative tension brane at an orbifold fixed point ($y=0$) in an infinite fifth dimension (the TeV brane).  Two additional positive tension branes are added at the points $y=\pm r$ (the Planck branes).  It is important to point out that unlike the theories proposed in \cite{LR} and \cite{GRS}, we place the TeV brane at the orbifold fixed point which (as we will discuss later) stabilizes the Radion mode (see \cite{pilo}).  The $Z_2$ symmetry demands that the tensions of the two additional Planck branes be equal.  The action takes the form.
\begin{equation}
    S=\int d^5x \sqrt{-g^{(5)}}\left[2 M_{pl}^{(5)\,3} R-\Lambda_b - \sqrt{-g^{(4)}}V_1\delta (y) - \sqrt{-g^{(4)}}V_2\left\{\delta (y+r)+\delta (y-r)\right\}\right].
\end{equation}
If we assume four-dimensional Poincare invariance, the metric is given by
\begin{equation}
    ds^2=g_{MN}\,dx^M dx^N
\end{equation}
with
\begin{equation}
    g_{MN}(x^\mu, y)= \begin{pmatrix}
            -e^{-A(y)} & 0 & 0 & 0 & 0 \\
            0 & e^{-A(y)} & 0 & 0 & 0 \\
            0 & 0 & e^{-A(y)} &  & 0 \\
            0 & 0 & 0 & e^{-A(y)} & 0 \\
            0 & 0 & 0 & 0 & 1
          \end{pmatrix}
\end{equation}
and

\begin{equation}\label{eq:ymetric}
    A(y)=\left\{ \begin{array}{ll}
                    -2 k_1 |y| & \mbox{if $0 \leq |y| \leq r$}\\
                    2 k_2 |y| - 2(k_1+k_2)r & \mbox{if $|y| > r$}.\end{array} \right.
\end{equation}

As in \cite{RSI}, the assumption of four-dimensional Poincare invariance leads one to derive the tension of the TeV brane located at $y=0$ to be $V_1=-24 M_{pl}^{(5)\,3}k_1$ and the cosmological constant between the Planck and TeV branes is $\Lambda_1=-24 M_{pl}^{(5)} k_1^2$.  Likewise, the tension on the Planck brane located at $y=r$ is found to be $V_2=24 M_{pl}^{(5)\,3}(k_1+k_2)$ and the cosmological constant outside the Planck branes is $\Lambda_2=-24 M_{pl}^{(5)} k_2^2$.
It is useful to transform the metric to manifestly conformally flat coordinates, where Einstein's equations take a simpler form.  In these coordinates, the metric takes the form
\begin{equation}\label{eq:zmetric}
    g_{MN}(x^\mu, z)= e^{-A(z)}diag(-1,1,1,1,1)
\end{equation}
where
\begin{equation}
    e^{-A(z)}=\left\{ \begin{array}{ll}
                    \frac{1}{(-k_1|z|+1)^2} & \mbox{if $z \leq z_b$}\\
                    \frac{1}{(k_2|z|+C)^2} & \mbox{if $z > z_b$}.\end{array} \right.
\end{equation}
Now the Planck branes are located at $z_b=\pm(1-e^{-k_1/r})/k_1$ and the constant $C=-k_2/k_1+\exp[-k_1 r](1+k_2/k_1)$ is chosen such that $z_b$ is the same for the two slices of AdS space.

\subsection{Kaluza-Klein Modes}

For now we will just consider the spin-2 fluctuation of the metric.  The scalar mode (radion) will be discussed in the following section.  Consider a pertubation of the form
\begin{equation}
    ds^2=e^{-A(z)}\left(dx^\mu dx^\nu(\eta_{\mu\nu}+h_{\mu\nu}(x,z))+dz^2\right).
\end{equation}
The transverse traceless solution can be written as $h_{\mu\nu}(x,z)=e^{3 A(z)/4}\tilde{h}_{\mu\nu}(x)\psi(z)$ where $\Box_4\, \tilde{h}_{\mu\nu}(x)=m^2 \tilde{h}_{\mu\nu}(x)$ and
\begin{equation}\label{eq:GravityDE}
    \left[-\partial_z^2+V(z)\right]\psi(z)=m^2\psi(z).
\end{equation}
The potential $V(z)$ is found to be \cite{RSII}
\begin{eqnarray*}
    V(z)&=&\frac{9}{16}(\partial_z A(z))^2-\frac{3}{4}\partial_z^2 A(z)\\
        &=&\left\{ \begin{array}{ll}
                    \frac{15k_1^2}{4(-k_1|z|+1)^2} & \mbox{if $|z| \leq zb$}\\
                    \frac{15k_2^2}{4(k_2|z|+C)^2} & \mbox{if $|z| > zb$}\end{array}\right\}
       +\frac{3k_1}{(-k_1+1)}\delta(z)\\
       & &-\frac{3}{2}
                   \left(\frac{k_1}{(-k_1|z|+1)}+\frac{k_2}{(k_2|z|+C)}\right)
                   \left(\delta(z-z_b)+\delta(z+z_b)\right).
\end{eqnarray*}
As usual, since the equation of motion for the Kaluza-Klein modes can be written in the form $\hat{Q}^\dag \hat{Q}\,
\psi(z)=m^2\,\psi(z)$ with $\hat{Q}=\partial_z+(3/4) A'(z)$, there is a zero mode solution that satisfies $\hat{Q}\,\psi_0(z)=0$:
\begin{equation}
    \psi_0(z)=N \exp[-\frac{3}{4}A(z)].
\end{equation}
N is found by normalization: $N=\left[\int \exp[-3/2 A(z)]dz\right]^{-1/2}$.

\noindent The higher KK modes are found by solving equation [\ref{eq:GravityDE}] subject to the following boundary conditions and normalization:
\begin{myindentpar}{2cm}
1) $\psi_m(z)$ is continuous at the Planck branes ($z=\pm z_b$).\\
2) $\psi_m'(z)$ is discontinuous at:
    \begin{myindentpar}{1cm}
    a) the TeV brane: $\Delta(\psi_m'(z))|_{z=0}=3k_1 \psi_m(0)$.\\
    b) the Planck branes: $\Delta(\psi_m'(z))|_{z=\pm z_b}=-\frac{3}{2}\left(
        \frac{k_1}{-k_1|z_b|+1}+\frac{k_2}{k_2|z_b|+C}\right) \psi_m(\pm z_b)$.
    \end{myindentpar}
3) $\psi_m(z)$ approaches a normalized plane wave solution for very large z.\\
\end{myindentpar}
The solution is
\begin{equation}
    \psi_m(z)=\left\{ \begin{array}{ll}
                    (-|z|+1/k_1)^{1/2}\left[a_m Y_2 (m(-|z|+1/k_1))+b_m J_2(m(-|z|+1/k_1))\right] & \mbox{if $|z| \leq zb$}\\
                    (|z|+C/k_2)^{1/2}\left[a'_m Y_2 (m(|z|+C/k_2))+b'_m J_2(m(|z|+C/k_2))\right] & \mbox{if $|z| > zb$}.\end{array} \right.
\end{equation}
The boundary conditions and normalization give the following relationships among the coefficients:
\begin{equation}
a_m Y_2(me^{-k_1 r}/k_1)+b_m J_2(me^{-k_1r}/k_1)=\\
     \left(\frac{k_1}{k_2}\right)^{1/2}\left[a'_m Y_2(me^{-k_1 r}/k_2)+b'_m J_2(me^{-k_1 r}/k_2)\right]
\end{equation}
\begin{equation}
a_mY_1(m/k_1)+b_mJ_1(m/k_1)=0
\end{equation}
\begin{equation}
a_m Y_1(me^{-k_1 r}/k_1)+b_m J_1(me^{-k_1r}/k_1)=\\
        \left(\frac{k_1}{k_2}\right)^{1/2}\left[a'_m Y_1(me^{-k_1 r}/k_2)+b'_m J_1(me^{-k_1 r}/k_2)\right]
\end{equation}
\begin{equation}
a_m^{'2}+b_m^{'2}=m.
\end{equation}


\begin{figure}[h]
\centering
\epsfig{figure=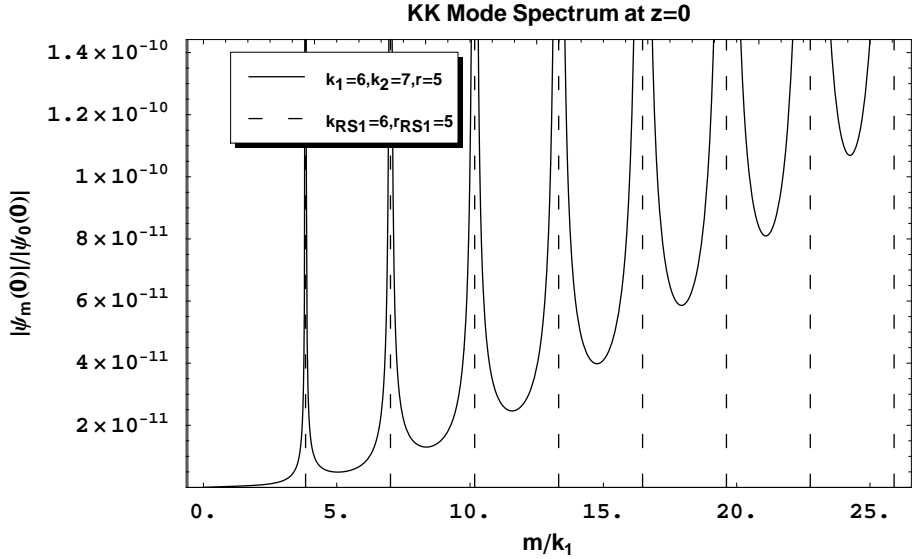,width=5.0in}
\caption{Mass Spectrum for both the Hybrid RS (solid) and RS1 (dashed) models.  The Hybrid RS model's spectrum was normalized by the zero mode's value at z=0.}
\label{fig:spectrumplot}
\end{figure}

\noindent Unlike the RS1 model, there is a continuous spectrum of graviton modes (all $m>0$ are allowed).  The RS1 spectrum is discrete and given by $m_n = k_1\,x_{n}$, where $x_n$ denotes the zeros of $J_1(x)$ \cite{csaki}\footnote{Since we have normalized the metric to be 1 at the TeV brane instead of the Planck brane as done in RS1 \cite{RSI}, our spectrum is multiplied by $\exp[k_1 r]$ as compared to the solution found in \cite{csaki}} .  In Fig. \ref{fig:spectrumplot} we compare the Hybrid RS KK spectrum to that of RS1.  We have chosen order one parameters such that $k_1 r=30$.  The resonances in the spectrum correspond nicely to the discrete spectrum found in RS1.  Since the modes are suppressed compared to the zero mode, the corrections to Newton's Law are small:
\begin{eqnarray}
    V(\bar{x},z=0,\bar{x}',z'=0)&=&\frac{1}{2 M_{pl}^3}\frac{|\psi_0(0)|^2}{|\bar{x}-\bar{x}'|}+\int_0^\infty\frac{1}{2 M_{pl}^3}
    \frac{|\psi_m(0)|^2 e^{-m|\bar{x}-\bar{x}'|}}{|\bar{x}-\bar{x}'|}dm\\
    &\sim&\frac{1}{2 M_{pl}^3}\frac{|\psi_0(0)|^2}{|\bar{x}-\bar{x}'|}\left(1+\int_0^\infty e^{-m|\bar{x}-\bar{x}'|}\frac{|\psi_m(0)|^2}{|\psi_0(0)|^2 }dm\right).
\end{eqnarray}
\subsection{Radion Stabilization}
As mentioned above, placing the TeV brane at the orbifold fixed point will allow the radion mode to be stabilized. To see this we need to include the spin-0 fluctuation of the 5 dimensional graviton.  The proper way to include this mode was discussed in \cite{pilo} and \cite{CGR}.  It was found that the metric can be written in such a way that the spin-2 calculation goes through as done above and is decoupled from the spin-0 radion mode ($f(x)$).  For the metric given in Equation \ref{eq:ymetric}, Pilo et al. \cite{pilo} found that the effective four dimensional lagrangian contains the term
\begin{equation}
    {\cal L} \supset \frac{24 M_{pl}^{(5)\,3}}{2 k_1}\left(1-e^{-2 k_1 r}\frac{k_2}{k_2+k_1} \right)
    \int \sqrt{-g}d^4x f\square f.
\end{equation}
Since the kinetic term is always positive in our model, positivity of energy is not violated.  The radion can be stabilized using a mechanism like the one introduced by Goldberger and Wise \cite{Goldberger}.

\section{Higgsless Symmetry Breaking in the Hybrid Model}\label{sec:gauge}

In this section we will put $\text{SU(2)}_L \times \text{SU(2)}_R
\times \text{U(1)}_{B-L}$ gauge fields in the bulk.  The metric is given by (\ref{eq:zmetric}) (see Fig. \ref{fig:higgslessmetric}).  However, unlike before, in this section we cut off the infinite extra dimension in order to make the massless mode normalizable \footnote{We will now use $r_1$ instead of r to denote the distance of the first brane to the origin. Also we will only consider half of the space for most of the discussion since the other half is obtained by orbifolding about the origin}.  This is
accomplished by adding a negative tension brane at an orbifold fixed point: $y=(r_1+r_2)$ (or $z=z_{b2} = 1/k_2(e^{(k_2 r_2 - k_1 r_1)}-k_2/k_1(e^{-k_1 r_1}-1+k_1/k_2 e^{-k_1 r_1}))$ in z-coordinates).
The 5D action for this model is:
\begin{equation}
S = \int d^4x \int dz \sqrt{-g^{(5)}}
\left[-\frac{1}{4}R^{a}_{MN}R^{aMN}-\frac{1}{4}L^{a}_{MN}L^{aMN}-\frac{1}{4}B_{MN}B^{MN}\right]\label{5Daction}
\end{equation}
where $R^{a}_{MN}$, $L^{a}_{MN}$, and $B_{MN}$ are the
$\text{SU(2)}_L$, $\text{SU(2)}_R$, and $ \text{U(1)}_{B-L}$ field
strengths.

Using the same procedure as \cite{Higgsless}, we chose to work in
unitary gauge where all KK modes of the fields $L_5^a, R_5^a, B_5$
are unphysical.  Boundary conditions were imposed to break the
$\text{SU(2)}_L \times \text{SU(2)}_R \times \text{U(1)}_{B-L}$ symmetry to the
Standard Model at $z=z_{b2}$ and to $\text{SU(2)}_D \times \text{U(1)}_{B-L}$ at $z =0$. The boundary conditions are:
\begin{eqnarray}
z=0:&&\left\{ \begin{array}{l}\partial_z(L_{\mu}^a+R^a_{\mu}) =
0,\,\, L_{\mu}^a -R^a_{\mu}=0,\,\, \partial_z B_{\mu} =0,\\
L_5^a+R_5^a = 0,\,\, \partial_z(L_5^a-R_5^a)=0,\,\, B_5 = 0
\end{array}\right.\\
z=z_{b2}:&&\left\{ \begin{array}{l}\partial_z
L^a_{\mu}=0,\,\,R^{1,2}_{\mu}=0\\ \partial_z (g_5
B_{\mu}+\tilde{g_5}R^3_{\mu})= 0,\,\, \tilde{g_5}B_{\mu}-g_5
R^{3}_{\mu} = 0,\\ L_{5}^{a}=0, \,\,R^{a}_{5}=0,\,\,B_5=0
\end{array}\right.
\end{eqnarray}
where $g_5$ and $\tilde{g_5}$ are the 5D gauge
coupling for $\text{SU(2)}_{L,R}$ and $\text{U(1)}_{B-L}$
respectively. In addition to the
boundary conditions we imposed continuity for the wave function
at $z = z_b$. The bulk equation of motion for the gauge fields is
\begin{equation}
\left[\partial_{z'}^2 -\frac{1}{z'}\partial_{z'}+\frac{q^2}{k_{1,2}^2}\right]\psi(z') = 0 \label{eqm}
\end{equation}
where $z'= -k_1 z +1$ or $k_2 z +C $  for $0 \leq z \leq z_b$ and $ z_b \leq z \leq z_{b2}$ respectively.  The solution to this equation is given by
\begin{equation}\label{wavefunction}
\psi^{d}_i = \left \{
\begin{array}{l}(-k_{1}z+1)\left(a_i^d \,J_1(q_i(-z+1/k_1))+b_i^d \,Y_1(q_i(-z+1/k_1))\right),\,\,\, 0\leq z \leq z_b\\
(k_{2}z+C)\left(a^{'\,d}_i \,J_1(q_i(z+C/k_2))+b^{'\,d}_i \, Y_1(q_i(z+C/k_2))\right),\,\,\,
z_b \leq z \leq z_{b2}\end{array}\right.
\end{equation}
where $d$ labels the corresponding gauge bosons ($W_{\pm}$, $L3$, $B$, $R3$).  Following \cite{Higgsless}, we expand the fields in their Kaluza-Klein modes as follows:
\begin{eqnarray}
B_\mu (x,z) &=& \frac{1}{\tilde{g_5}}a_{0}\gamma(x)+\sum_{j=1}^{\infty}\psi_{j}^{B}(z)Z_\mu^j (x)\\
L^3_\mu (x,z) &=& \frac{1}{g_5}a_{0}\gamma(x)+\sum_{j=1}^{\infty}\psi_{j}^{L3}(z)Z_\mu^j (x)\\
R^3_\mu (x,z) &=& \frac{1}{g_5}a_{0}\gamma(x)+\sum_{j=1}^{\infty}\psi_{j}^{R3}(z)Z_\mu^j (x)\\
L^\pm_\mu (x,z) &=& \sum_{j=1}^{\infty}\psi_{j}^{L_\pm}(z)W_\mu^{j\pm} (x)\\
R^\pm_\mu (x,z) &=& \sum_{j=1}^{\infty}\psi_{j}^{R_\pm}(z)W_\mu^{j\pm} (x)
\end{eqnarray}

\subsection{Oblique Corrections}
In order to calculate the electroweak corrections in our model we ensure that all corrections are oblique.  This is done by adjusting the coupling of the fermions localized at $z=z_{b2}$ so that the zero mode couplings are equal to the SM couplings at tree level.  For our model the relations are
\begin{eqnarray}\label{eqncouplings}
-\frac{\tilde{g_5}\psi_{1}^{(B)}(z_{b2})}{g_{5}\psi_{1}^{(L3)}(z_{b2})}&=&\frac{g'^2}{g^2}\\
g_{5}\psi_{1}^{(L\pm)}(z_{b2})&=& g \label{eqnW}\\
g_{5}\psi_{1}^{(L3)}(z_{b2})&=& g\cos \theta_W \label{eqnZ}
\end{eqnarray}
For the photon kinetic term, we canonically normalize it as follows:
\begin{eqnarray}\label{eqnphoton}
Z_{\gamma} &=&\left(\left( a_0/\tilde{g_5}\right)^2+\left( a_0/g_5\right)^2\right) I=1\\
I&=&\int_{-z_{b2}}^{z_{b2}}e^{-A(z)/2} dz.
\end{eqnarray}
Equations (\ref{eqnW}) and (\ref{eqnZ}) are used to determine the correct normalization for the W and Z wavefunctions.

\begin{figure}[h]
\centering
\epsfig{figure=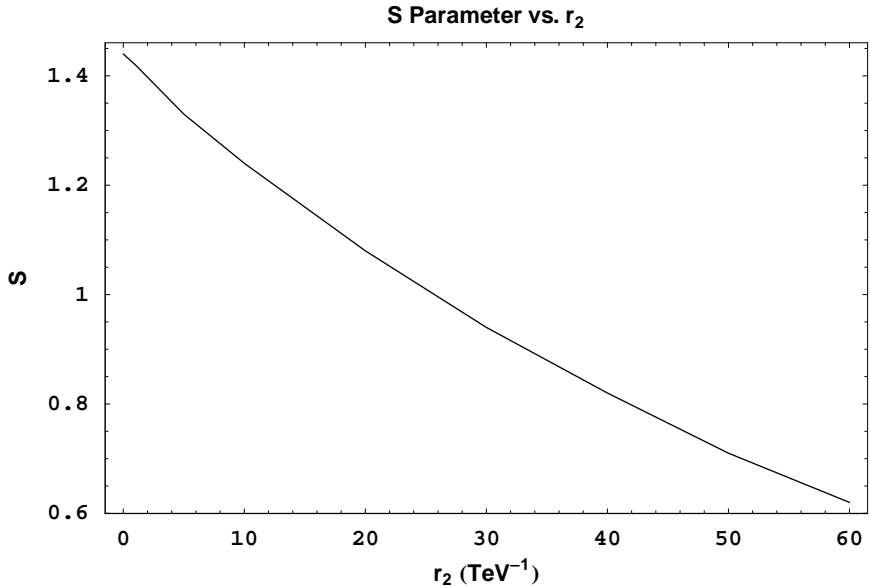,width=5.0in}
\caption{Plot of the S parameter as a function of $r_2$.}
\label{fig:Splot}
\end{figure}

Given the gauge field's wavefunctions, we calculated the oblique corrections using the relations between the vacuum polarization and the wavefunction renormalization: $Z_{\gamma} = 1-\Pi'_{QQ}$, $Z_{W} = 1-g^2\Pi'_{11}$, and $Z_{Z} = 1-(g^2+g'^2)\Pi'_{33}$ \cite{Peskin-Takeuchi}. The wavefunction renormalizations are give by
\begin{eqnarray}\label{eq:Z}
Z_{W} &=&\int_{-z_{b2}}^{z_{b2}}\left[\psi^W\right]^2 e^{-A(z)/2} dz=\int_{-z_{b2}}^{z_{b2}}\left(\left[\psi^{L_+}\right]^2+\left[\psi^{R_+}\right]^2\right)e^{-A(z)/2} dz\\
Z_{Z} &=&\int_{-z_{b2}}^{z_{b2}}\left[\psi^Z\right]^2e^{-A(z)/2} dz=\int_{-z_{b2}}^{z_{b2}}\left(\left[\psi^{L3}\right]^2+\left[\psi^{R3}\right]^2+\left[\psi^{B}\right]^2\right)e^{-A(z)/2} dz,
\end{eqnarray}
and the zero momentum vacuum polarizations are
\begin{eqnarray}
\Pi_{11}(0) &=& \frac{1}{g^2}\int_{-z_{b2}}^{z_{b2}}\left(\left[\partial_z\psi^{L_+}\right]^2+\left[\partial_z\psi^{R_+}\right]^2\right)e^{-A(z)/2} dz\\
\Pi_{33}(0) &=& \frac{1}{g^2+g'^2}\int_{-z_{b2}}^{z_{b2}}\left(\left[\partial_z\psi^{L3}\right]^2+\left[\partial_z\psi^{R3}\right]^2
+\left[\partial_z\psi^{B}\right]^2\right)e^{-A(z)/2} dz.\\
\end{eqnarray}
The Peskin-Takeuchi oblique corrections as a function of vacuum polarization are defined as \cite{Peskin-Takeuchi} :
\begin{eqnarray}\label{eq:STU}
S&=& 16\pi(\Pi'_{33}-\Pi'_{3Q})\\
T&=&\frac{4\pi}{\sin^2\theta_W\cos^2\theta_W M_Z^2}(\Pi_{11}(0)-\Pi_{33}(0))\\
U&=&16\pi(\Pi'_{11}-\Pi'_{33})
\end{eqnarray}
Since we are only considering the tree level corrections, $\Pi'_{3Q}=0$.
As an input to our model, we use the values of the SM electroweak parameters at the Z-pole: $M_W = 80.045$ GeV, $\sin^2\theta_W=0.231$, and $\alpha = 127.9$. We also assume $k_1 r_1=30$.  In the limit $r_2 \rightarrow 0$, $M_W$ sets the size of the extra dimension to be $r_1=68.5$ TeV$^{-1}$.  Since this is the limit of the standard higgsless model, we find $T=U=0$ and $S\sim 6 \pi/(g^2 (k_1 r_1))\sim 1.4$ as in \cite{Higgsless}.  Since we are only interested in showing that the $S$ parameter decreases while preserving $T\sim0$ and unitarity, we do not do a complete survey of the parameter space.  For our analysis we set $k_2$ to be equal to the value of $k_1$ in the $r_2 \rightarrow 0$ limit. As we increase $r_2$, we find $r_1$ decreases in order to produce the proper $M_W$.  Fig. \ref{fig:Splot} shows the behavior of the S parameter as we increase $r_2$.  We find the $S$ parameter decreases. For $r_2=60$ we also checked that the lightest $W$ and $Z$ excitations are less than 1800 GeV and therefore unitarity is preserved \cite{Higgsless}. This provides another mechanism for lowering the $S$ parmeter in addition to including brane kinetic terms \cite{Higgsless} and bulk fermions \cite{Cacciapaglia}.

\section{Conclusions}\label{sec:conclusions}
In the first section, we presented a model that is a hybrid between RSI and RSII.  The model has a negative tension brane located at an orbifold fixed point ($y=0$) and two identical positive tension branes located at $y=\pm r$.  The fifth dimension extends to infinity as in RSII, however the presence of the positive tension branes produces graviton resonances which coincide with the discrete RSI spectrum.  This model is attractive since it both solves the hierarchy problem and produces a continuum of KK graviton modes.  As in both the RSI and RSII models, four dimensional gravity can be recovered.  Stability of our model is ensured by placing the negative tension brane at an orbifold fixed point.

In the second section of the paper, negative tension branes were brought in from infinity to cut off the space at an orbifold fixed point.  We included SU$(2)_{\rm{L}} $x SU$(2)_{\rm{R}} $x U$(1)_{\rm{B-L}}$ fields in the bulk and broke to the Standard Model on the far brane.  The distances between the branes are scaled as to produce the correct W mass. As in standard higgsless electroweak symmetry breaking models, a large $S$ parameter along with vanishing $T$ and $U$ parameters were found when the second slice of our space was shrunk to zero.  As the second slice of our space was increased, the $S$ parameter was lowered while corrections to both $T$ and $U$ remained suppressed.  We also find the lightest $W$ and $Z$ excitations stayed below 1800 GeV and therefore preserve unitarity. In conjunction with using brane kinetic terms and placing fermions in the bulk, this could be used as a useful mechanism for lowering the $S$ parameter.

Future work on these models could include trying to incorporate both higgsless electroweak symmetry breaking and solutions to the hierarchy problem into a single model.  It would also be interesting to explore how this model compares to other known mechanisms used to lower the $S$ parameter.



\section*{Acknowledgments}
We thank Josh Erlich for his invaluable comments during the preparation this paper.  We would also like to thank Marc Sher and Chris Carone for useful discussion and comments.  B.G. and J.T. were supported by NSF grant PHY-0757481.

\end{document}